\documentclass[twocolumn]{pasj01}
\usepackage{amsmath}
 \usepackage{url}
\usepackage[switch,mathlines]{lineno}

\newcommand{\yr}{\mathrm{yr}}

\newcommand{\erg}{\mathrm{erg}}

\newcommand{\s}{\mathrm{s}}

\newcommand{\cm}{\mathrm{cm}}

\newcommand{\iras}{IRAS 00500+6713}

\newcommand{\red}{\textcolor{black}}

\usepackage{CJKutf8}

\usepackage{chngcntr}
\begin{document} 
\Received{}
\Accepted{}

\title{Delayed Wind Onset in Pa 30, the Remnant of Type Iax SN 1181}

\author{Takatoshi \textsc{Ko}\altaffilmark{1,2}
\email{ko-takatoshi@resceu.s.u-tokyo.ac.jp}}
\altaffiltext{1}{Research Center for the Early Universe (RESCEU), School of Science, The University of Tokyo, 7-3-1 Hongo, Bunkyo-ku, Tokyo 113-0033, Japan}
\altaffiltext{2}{Department of Astronomy, School of Science, The University of Tokyo, 7-3-1 Hongo, Bunkyo-ku, Tokyo 113-0033, Japan}

\author{Ryosuke \textsc{Hirai}\altaffilmark{3,4,5}
}
\altaffiltext{3}{Astrophysical Big Bang Laboratory, Pioneering Research Institute, RIKEN, 2-1 Hirosawa, Wako, Saitama 351-0198, Japan}
\altaffiltext{4}{School of Physics and Astronomy, Monash University, Clayton, Victoria 3800, Australia}
\altaffiltext{5}{OzGrav: The ARC Centre of Excellence for Gravitational Wave Discovery, Australia}

\author{Taiga \textsc{Sasaoka}\altaffilmark{6}
}
\altaffiltext{6}{Institute of Astronomy, Graduate School of Science, The University of Tokyo, 2-21-1 Osawa, Mitaka, Tokyo 181-0015, Japan}

\author{Toshikazu \textsc{Shigeyama}\altaffilmark{1,2}
}
\maketitle

\begin{abstract}
Pa 30 is the recently identified remnant of the historical supernova SN 1181, likely a Type Iax event, and a nebula surrounding the central white dwarf launching a fast wind ($\sim10^9~\cm~\s^{-1}$) is observed in optical and infrared bands. X-ray observations show that this wind collides with the surrounding material and produces a termination shock, and the observed extent of the shock indicates that the wind started blowing centuries after 1181 A.D. rather than immediately after the SN explosion. We propose that the wind is triggered by delayed ignition of fallback carbon-rich material on the WD surface and investigate the conditions that reproduce such delayed ignition. We show that producing delays of several centuries requires a relatively hot post-explosion WD core with a temperature $T_c \simeq 6\times10^8~\mathrm{K}$. This supports the pure-deflagration progenitor scenario for Type Iax SN 1181, which implies the presence of a He star companion inside Pa~30; we also discuss why such a potential He star has not been detected and its prospects for discovery by future observations.
\end{abstract}

\clearpage 

\section{Introduction} \label{sec:Intro}
Since ancient times, supernovae (SNe) have been recorded in historical documents across the world, occurring roughly once every few centuries at distances close enough to be visible to the naked eye. Their remnants (historical SNRs) have therefore been studied extensively. SN~1006 is the brightest historical event; SN~1054 is identified with the Crab Nebula, a prototypical pulsar-wind nebula; SN~1572 is known as Tycho’s SN; and SN~1604 as Kepler’s SN. Owing to their brightness, these remnants have been recognized and investigated since the 1920s–1930s (see \cite{2017hsn..book...37G}, for a review).

Among historical SNRs, the most recent identification concerns the SN recorded in 1181. \citet{2021ApJ...918L..33R} reported the infrared nebula Pa~30 (also \iras) roughly in the location of the historical SN, and from its expansion velocity ($\sim1000~\mathrm{km~s^{-1}}$) and size ($\sim1$~pc) inferred an age of $\sim1000$~yr, \red{consistent with the estimate by X-ray observation~\citep{Oskinova_et_al_20}}, making Pa~30 the leading candidate counterpart to the SN~1181 remnant. At its center lies a peculiar object that drives an optically thick wind with a terminal velocity $\sim1.5\times10^{4}~\mathrm{km~s^{-1}}$ and a mass-loss rate of $\sim10^{-6}~M_\odot~\mathrm{yr^{-1}}$, as inferred from optical spectra \citep{2019Natur.569..684G,Lykou2022}. The photospheric composition, combined with the wind velocity, suggests that an oxygen–neon (ONe) white dwarf (WD) \citep{2019Natur.569..684G,2019ApJ...887...39K,Oskinova_et_al_20,Lykou2022} exists inside.

Historical records indicate that SN 1181 reached a brightness comparable to Saturn, implying a peak absolute magnitude of roughly \red{-16 to -12.5~\citep{2021ApJ...918L..33R,2023MNRAS.523.3885S}}\footnote{These estimates are based on a Japanese record \textit{Azumakagami}, and is independently supported by an Arabic poem that describe "a new bright star" appears in 1181 to 1182 A.D. in Cassiopeia brighter than any fixed star in that constellation~\citep{2025AN....34670024F}; the accounts are mutually consistent.}. Pa~30 is also observed in X-rays with \textit{XMM-Newton} and \textit{Chandra} \citep{Oskinova_et_al_20,2024ApJ...969..116K}, and two thermal components are detected. Modeling of its X-ray emission yields an ejecta mass of $0.2$–$0.5~M_\odot$ and an explosion energy of $\sim10^{48}~\mathrm{erg}$. These values are consistent with low-energy Type~Iax explosions and support the identification of Pa~30 as the SN~1181 remnant \citep{2024ApJ...969..116K}.

\begin{figure}

 \includegraphics[width=\linewidth]{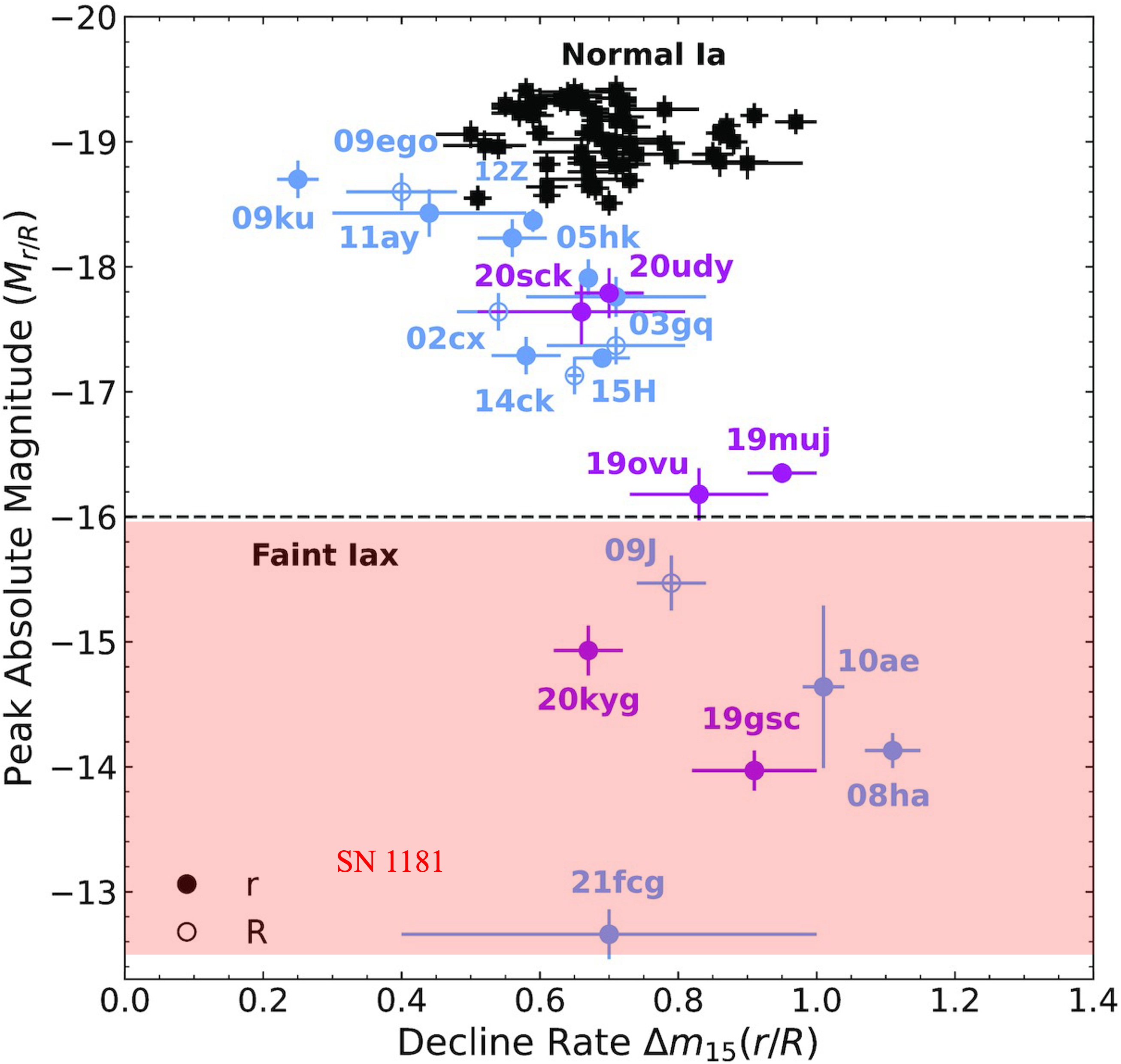}
\caption{Decline rate $\Delta m_{15}$ versus peak absolute magnitude $M$ for normal SNe~Ia and for the “bright” and “faint” subclasses of SNe~Iax (reproduced from Fig.~11 of \cite{2022MNRAS.511.2708S}; filled/open symbols denote $r$/$R$). The red region marks SN~1181; its values are not measured in $r/R$ but inferred from historical visual (naked-eye) estimates: a peak $M\simeq-12.5$ to \red{$-16$~mag~\citep{2021ApJ...918L..33R,2023MNRAS.523.3885S}.} The SN~1181 region is placed on the $r/R$ plane for comparison only.}

 \label{fig:SN1181mag}
\end{figure}

Type~Iax SNe are a subclass of thermonuclear SNe characterized by low peak luminosities (absolute magnitudes $\sim-13$ to $-18$), slow expansion velocities ($\sim6000$–$7000~\mathrm{km~s^{-1}}$ at maximum light), and the absence of the near-infrared secondary maximum seen in normal SNe~Ia (see \cite{2017hsn..book..375J}, for a review). They are often categorized into “bright’’ and “faint’’ groups (see Fig.~\ref{fig:SN1181mag}). The leading theoretical model invokes a pure-deflagration in a WD: the flame does not transition to a detonation and propagates subsonically, yielding lower explosion energies and typically a bound remnant. Pure-deflagration models of CO WDs reproduce many observed properties of the bright Iax events \citep{2012ApJ...761L..23J,2013MNRAS.429.2287K,2014MNRAS.438.1762F,2022A&A...658A.179L}, while hybrid CONe WDs can account for bright events and some faint ones \citep{2015MNRAS.450.3045K,2023ApJ...959..112F}. In this framework, the deflagration is triggered by accretion of He-rich material from a He star. \textit{HST} pre-explosion imaging of the type Iax SN~2012Z revealed a He-rich companion candidate \citep{2014Natur.512...54M}, \red{and early spectra of some SNe~Iax show He emission lines that likely originate in the circumstellar environment \citep{2019MNRAS.487.2538J}.} 
However, reproducing the faintest SNe~Iax within pure-deflagration models often requires very small ejecta masses, in tension with SN~2021fcg—the faintest Iax yet observed—whose inferred ejecta mass is comparable to that of brighter events~\citep{2021ApJ...921L...6K}. 

An alternative channel is an ONe + CO WD merger in which only the CO component detonates, leaving the ONe WD as the remnant \citep{2018ApJ...869..140K,Oskinova_et_al_20}; this pathway generally does not reproduce light curves and spectra of SNe Iax as well as pure-deflagration models. Moreover, recently identified intermediate-luminosity events such as SN 2019muj~\citep{2021MNRAS.501.1078B,2021PASJ...73.1295K,2022ApJ...941...15M}, SN 2022xlp~\citep{2025A&A...703A..64B}, SN 2024pxl~\citep{2025arXiv250502943S,2025ApJ...988..209H,2025ApJ...989L..33K} and SN 2025qe~\citep{2025MNRAS.543.3731M} show features overlapping with those of both the bright and faint groups, suggesting a continuous character rather than two discrete subclasses. Taken together, these considerations keep the progenitor systems of SNe~Iax under active discussion. Recent work further argues that in single-degenerate systems a deflagration–detonation transition may often fail, favoring Iax-like, low-energy outcomes over normal SNe~Ia \citep{2025arXiv250716907M}. Clarifying the progenitors of SNe~Iax is therefore central to the broader SN~Ia progenitor problem.

\red{Several recently studied Type~Iax SN remnants show late-time excess emission (e.g., SN~2012Z; \cite{2022ApJ...925..138M}). In some cases, their spectra/SEDs are better explained by a two-component model (e.g., SN~2014dt; \cite{2016MNRAS.461..433F}; SN~2012Z; \cite{2025arXiv250401063S}; and SN~2024pxl and SN~2024vjm; \cite{2025ApJ...989L..33K}). The two-component spectrum consists of a nebular component plus an additional blackbody-like continuum. The blackbody component is interpreted as arising from an optically thick wind launched from the bound remnant, as suggested for Pa~30. The wind could be powered by delayed radioactive decays of $^{56}$Ni and $^{56}$Co \citep{2017ApJ...834..180S}.}
 Their characteristic velocities ($\sim500$–$1000~\mathrm{km~s^{-1}}$) are, however, far below the extreme wind inferred for the central WD in Pa~30. To account for this exceptionally fast wind, \citet{2019ApJ...887...39K} proposed a model involving a rapidly rotating, strongly magnetized WD that accelerates a carbon-ash outflow where the main energy source is shell carbon burning; the evolutionary pathway to such a configuration has not been systematically explored. 

In the SN 1181 remnant, the fast wind from the central WD collides with the SN ejecta and produces the termination shock detected in X-rays~\citep{Oskinova_et_al_20,2024ApJ...969..116K}. The combination of the very fast wind and the compact shock radius ($\sim0.01$ pc) indicates that the fast wind did not begin immediately after the SN 1181 explosion but turned on with a delay. One-dimensional thin-shell modeling of the termination shock colliding with the SN 1181 ejecta requires the wind onset to be in the 1990s \citep{2024ApJ...969..116K}. 
This presence of the delayed wind suggests delayed carbon ignition on the WD surface. Given the low explosion energy of SN~1181, at the faint end of the Iax population (Fig.~\ref{fig:SN1181mag}), some fraction of the CO-rich ejecta is expected to remain bound and fall back onto the central ONe WD. Determining the conditions under which fallback re-ignites surface burning and drives the delayed wind is therefore crucial to elucidating the fast-wind mechanism and the nature of the surviving central WD, and to constrain the progenitor system of SN~1181.

 Prior studies of ONe WDs have chiefly examined He accretion and its thermonuclear consequences~(e.g.,~\cite{2017ApJ...843..151B,2024ApJ...975..186Z}), or merger-driven setups often deposit CO onto an ONe WD at super–Chandrasekhar total masses (e.g., $M_{\rm ONe}\!\approx\!1.20~M_\odot$ with a $1.05~M_\odot$ CO WD) and follow prompt explosive outcomes~(e.g.,~\cite{2023ApJ...944L..54W,2024ApJ...967L..45W}). These approaches do not address the case of a modest CO accretion onto a massive ONe WD and the following thermal relaxation. Therefore, in this paper, we simulate the time evolution after CO-rich accretion onto an ONe WD using the public stellar evolution code \texttt{MESA}~(Modules for Experiments in Stellar Astrophysics;~\cite{2011ApJS..192....3P,2013ApJS..208....4P,2015ApJS..220...15P,2018ApJS..234...34P,2019ApJS..243...10P,2023ApJS..265...15J}) in order to explain the delayed carbon ignition in Pa30.

This paper is structured as follows. In Section~\ref{sec:method}, the \texttt{MESA} setups for the initial WD models are described. In Section~\ref{sec:results}, we report the results of the carbon ignition timescales and the conditions enabling wind delays of $ 800$ yr. In Section~\ref{sec:discussion}, we discuss the implications for the SN~1181 progenitor system and the connection between existing accretion studies. We conclude in Section \ref{sec:conclusion}.

\section{Methods}\label{sec:method}
\begin{figure*}

 \includegraphics[width=\linewidth]{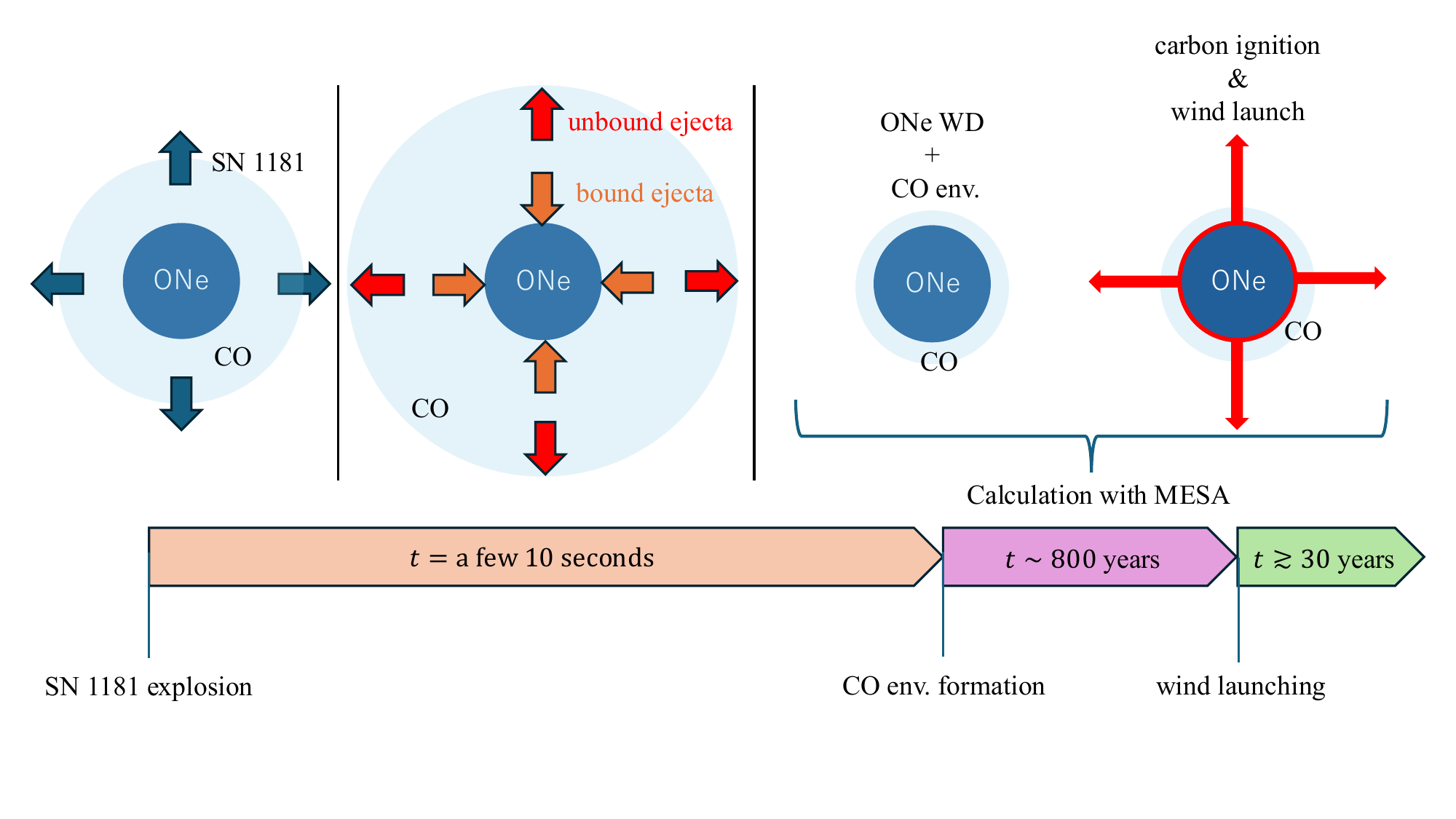}
\caption{Systematic diagram of the delayed-wind scenario examined in this work. In 1181 A.D. a Type Iax SN occurred. A fraction of the ejecta remained gravitationally bound and later fell back onto the central ONe WD. The fallback material formed an extended CO envelope on the WD surface, which eventually thermally contracts to reach higher temperatures and triggers carbon ignition, generating the energy necessary to launch the powerful wind. We model the evolution after the CO envelope has formed, up to the onset of surface carbon ignition.}
 \label{fig:schematic_pic}
\end{figure*}
\label{sec:Methods}
We investigate conditions for the ignition of fallback matter accreted onto the central WD inside Pa~30 to find the parameter space that allows for the delayed carbon ignition on timescales of $\sim800$ yr. As discussed in Sec~\ref{sec:Intro}, the pure-deflagration model remains the leading progenitor scenario for SNe Iax, while merger channels involving an ONe WD and CO WD have also been proposed. Because simulations across these scenarios indicate that SN Iax ejecta are CO-rich~(e.g.,~\cite{2014MNRAS.438.1762F,2018ApJ...869..140K}), we restrict our analysis to accretion of matter containing only CO for simplicity. In pure-deflagration scenarios, the bound remnant at the center may be a CO or ONe WD, whereas in the WD merger channel the central remnant is generally an ONe WD. For Pa~30, prior estimates place the WD mass at $\sim1.1$–$1.3~M_\odot$~(e.g.,~\cite{2019ApJ...887...39K,Lykou2022}), \red{which is more in line with typical masses of ONe WDs; CO WDs have masses around $\sim0.6~M_\odot$~(\cite{1990sse..book.....K} for a review.).}

Given the specific context of SN 1181, the explosion energy is estimated to be as low as $\sim10^{48}\,\mathrm{erg}$~\citep{2024ApJ...969..116K}, which is significantly smaller than the gravitational binding energy for a SN Iax ejecta of order $0.1~M_\odot$ at the surface of a typical ONe WD ($\sim10^{50}~\erg$). As a result, the bulk of the ejecta is expected to fall back and be accreted onto the WD on a timescale comparable to the dynamical timescale near the surface, which is on the order of seconds to tens of seconds. Since this timescale is much shorter than the post-accretion evolution timescales we focus on ($\sim1000$ years), we neglect the accretion process itself and instead construct an initial condition of an ONe WD already surrounded by a hydrostatic CO envelope (see Fig.~\ref{fig:schematic_pic}).

We construct the ONe WD structure with an overlying CO envelope using the public stellar evolution code \texttt{MESA}~(version r24.08.1;~\cite{2011ApJS..192....3P,2013ApJS..208....4P,2015ApJS..220...15P,2018ApJS..234...34P,2019ApJS..243...10P,2023ApJS..265...15J}) and follow its time evolution until the onset of carbon ignition. We adopt an ONe WD core to enable carbon shell burning, as accretion of CO-rich material onto a CO WD would not yield a shell-like burning layer. For simplicity, we assume a homogeneous composition with $X(^{16}\mathrm{O})=X(^{20}\mathrm{Ne})=0.5$. We first set up an ONe WD to serve as the accretor for subsequent accretion of CO-rich material. The WD is composed of a degenerate core and a partially degenerate envelope. Here, we define regions where the electron degeneracy parameter is $\eta>1$ as "degenerate".
The procedure for attaching the CO envelope to the ONe WD is detailed in Section~\ref{sec:attach_co}, and its subsequent evolution is presented in Section~\ref{sec:results}.

For control experiments, a set of five idealized models are prepared, as summarized in Table~\ref{tab:init}. Each of our models consist of two main zones; a near-isothermal ONe core and a CO envelope. The central ONe core is mostly degenerate and therefore effectively isothermal due to electron conductivity. The exact temperature depends on the details of the preceding explosion. Similarly, thermal properties of the CO envelope depends strongly on the details of the fallback process. Given the large uncertainties in both the explosion and fallback processes, we treat the thermal properties of each zone as free parameters. By performing our simulations over a grid of model parameters and comparing the results with the observed delay time, we can constrain the required thermal properties and provide insight into the explosion and fallback mechanisms. 
 For the mass, two different central WD masses ($M_{\mathrm{ONe}}$) of $1.1~M_\odot$ and $1.2~M_\odot$ are adopted, motivated by estimates of the mass of the WD in Pa~30 ($1.1$–$1.3~M_\odot$). To be consistent with the characteristic ejecta mass scale of SNe~Iax ($\mathcal{O}(0.1~M_\odot)$), total accreted masses ($M_{\mathrm{env,tot}}$) of $0.05~M_\odot$ and $0.10~M_\odot$ are explored. The accreted material is taken to be either pure carbon or a CO mixture. For the thermal state, we adopt ONe WDs with central temperatures $T_c=4\times10^8$ and $6\times10^8$~K, chosen to stay below the $\sim10^9$ K scale where degeneracy eases in ONe cores and hot enough to avoid numerical issues with colder WD models in \texttt{MESA}. 
As for the CO envelope, we adopt the radial extent of the outer CO envelope \(R_{\rm env}\) as a proxy for the thermal content of the envelope. For a given mass and composition, hotter (higher-entropy) envelopes are more extended. A broad but computationally feasible range of this extent is explored.

\begin{table*}
    \begin{center}

  \begin{tabular}{ccccccc}
    \hline
    &$M_{\mathrm{ONe}}~(M_\odot)$ & $T_c~(10^8~\mathrm{K})$ & $M_{\mathrm{env,tot}}~(M_\odot)$ & $M_{\mathrm{env,c}}~(M_\odot)$ & $M_{\mathrm{env,o}}~(M_\odot)$ & $R_{\mathrm{env}}~(10^8~\cm)$ \\
    \hline
   model 1&1.1 & 6 & 0.050 & 0.025 & 0.025 & 5.0--20 \\
    model 2&1.1 &4 &  0.050  & 0.025 & 0.025 & 5.0--20 \\
    
   model 3& 1.1 & 6 & 0.10 & 0.050 & 0.050 & 5.0--15 \\
   model 4&1.1 & 6 & 0.050 & 0.050 & 0 & 5.0--20 \\
    model 5&1.2 & 6 &  0.050  & 0.025 & 0.025 & 5.0--20 \\
    \hline
  \end{tabular}
  \end{center}
  \caption{Model grid adopted in this study. Columns list the central ONe WD mass $M_{\rm ONe}$, core temperature $T_c$, total CO envelope mass $M_{\rm env,tot}$, the carbon and oxygen components ($M_{\rm env,c}$, $M_{\rm env,o}$), and the envelope radius $R_{\rm env}$. For each model, multiple $R_{\rm env}$ values within the indicated range are explored; Model~1 serves as the fiducial setup.}
\label{tab:init}
\end{table*}

\subsection{Attaching CO Envelope onto ONe White Dwarf}
\label{sec:attach_co}
In this section, we describe how we attach the CO envelope to the ONe WD. A full treatment of SN 1181 ejecta fallback after SN 1181 depends on the SN ejecta energy distribution and would be complex and is not pursued here. For simplicity, we artificially reproduce the ONe WD with CO envelope that approximates the post-accretion state.
The accretion setup follows the \texttt{wd\_acc\_small\_dm} case in the \texttt{MESA} \texttt{test\_suite}, with minor modifications for our application. The mass and abundance of the accreted material are summarized in Table~\ref{tab:init}. Here, accretion serves merely as a means to append the outer layer, and we do not explicitly follow the accretion process itself; therefore the accretion rate is not relevant. In reality, the fallback material would settle on a timescale of a few~$\times 10$~seconds, sufficiently shorter than the cooling time of the ONe WD, so the ONe WD should not have cooled even after fallback. However, realizing such a rapid deposition is difficult within \texttt{MESA}, and thus, to reproduce that situation, we keep the central WD effectively isothermal while accretion proceeds. To maintain the isothermal core during the CO accretion phase, we inject artificial heat into zones within the core which have sub-target temperatures ($T<T_c$) using the \texttt{extra\_heat} control in \texttt{MESA}. The nuclear reaction network is disabled during this stage.
Convective mixing is suppressed by setting \texttt{mix\_factor} to zero. The atmospheric boundary condition follows the original \texttt{wd\_acc\_small\_dm} setup, adopting an Eddington $T$–$\tau$ relation with iterated opacities.

Once the CO envelope with the specified mass is attached, we then apply a uniform, per-unit-mass heating at a constant specific rate ($\mathrm{erg}~\mathrm{g}^{-1}~\mathrm{s}^{-1}$) throughout the envelope to adjust the thermal profile so that it achieves the target $R_{\rm env}$. Artificial heating of the CO envelope and of the ONe WD is applied independently. We regulate the envelope radius to the target envelope radius by toggling the artificial envelope heating on when the current radius is smaller than the target and off when it is larger. We then evolve the system until a quasi-stationary state is reached near this switching boundary and adopt that controlled “equilibrium” structure as the initial conditions for our simulations. Note, the artificial heating is a numerical device and not a physical energy source.

\begin{figure}

 \includegraphics[width=\linewidth]{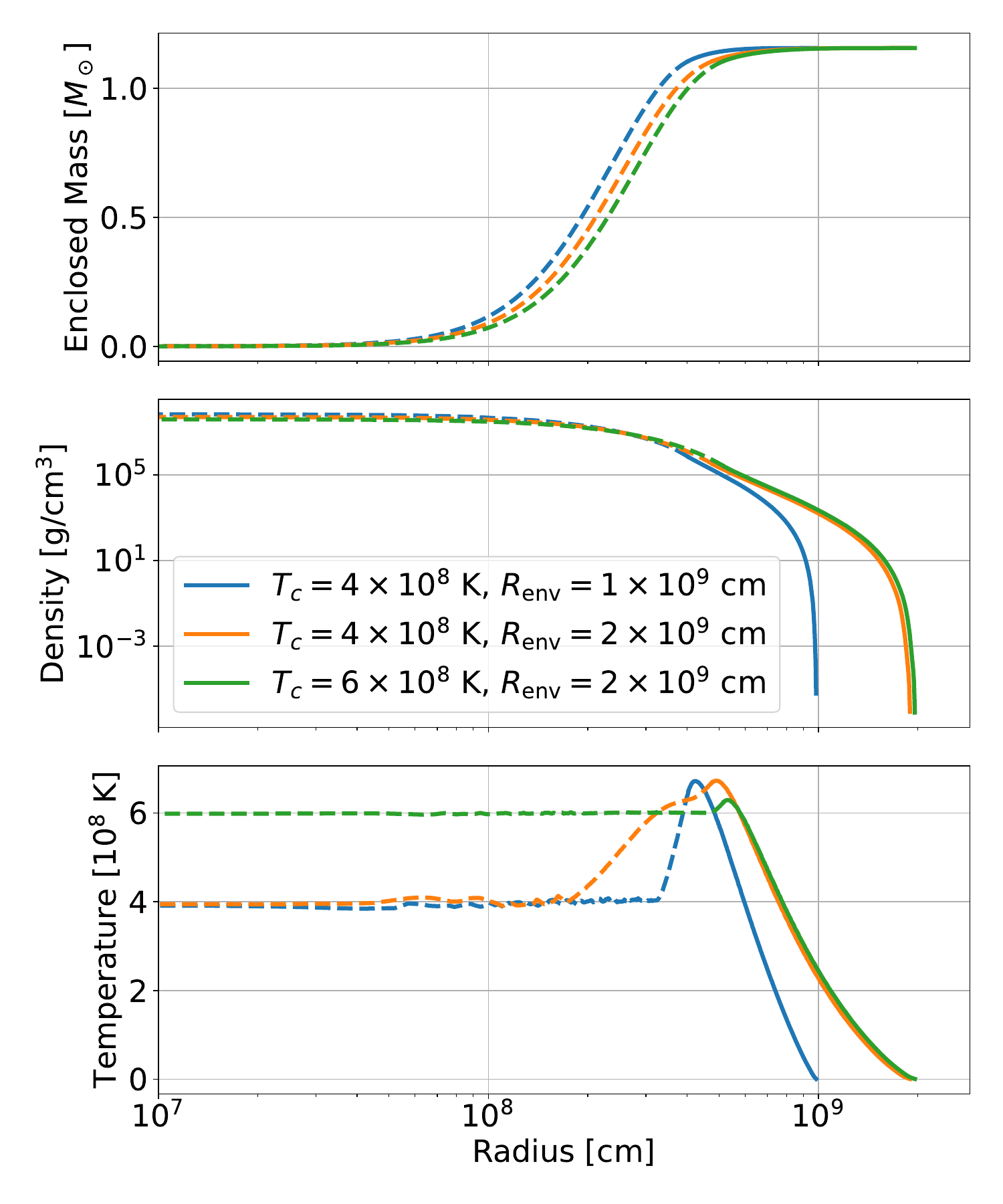}
 \caption{Initial radial profiles of density, temperature, and enclosed mass for representative models (\(M_{\rm WD}=1.1\,M_\odot\), $M_{\rm env,tot} = 0.05~M_\odot$, \(T_c=4\times10^8\) or \(6\times10^8~\mathrm{K}\), \(R_{\rm env}=1.0\times10^9\) or \(2.0\times10^9\,\mathrm{cm}\)). Dashed curves indicate the ONe composition of the central WD, and solid curves indicate the CO envelope.}
 \label{fig:init_features}
\end{figure}

Representative initial profiles for selected models ($M_{\rm WD}=1.1\,M_\odot$, $M_{\rm env,tot} = 0.050~M_\odot$, $T_c=4\times10^8$ or $6\times10^8$~K, $R_{\rm env}=1.0\times10^9$ or $2.0\times10^9$ cm) are shown in Figure~\ref{fig:init_features}. Dashed portions of the curves trace the physical quantities of the central ONe WD, while the solid portions correspond to the CO envelope. Electrons in the outer ONe layers are weakly degenerate and the profiles connect smoothly to the bottom of the CO envelope.

\section{Results}\label{sec:results}
\subsection{Time Evolution until carbon ignition}\label{sec:evolve}
We follow the time evolution of the models constructed in Section~\ref{sec:attach_co} up to the onset of carbon ignition. During these calculations, we turn off the artificial heat injection and track the evolution driven by neutrino cooling in the ONe WD core and radiative heat transport within the envelope. The numerical setup largely follows the configuration \texttt{inlist\_settle\_envelope} in the \texttt{MESA} \texttt{test\_suite} case \texttt{make\_o\_ne\_wd}. The details are available on Zenodo under an open-source Creative Commons Attribution license\footnote{\url{doi.org/10.5281/zenodo.17569071}}.

\begin{figure*}

 \includegraphics[width=\linewidth]{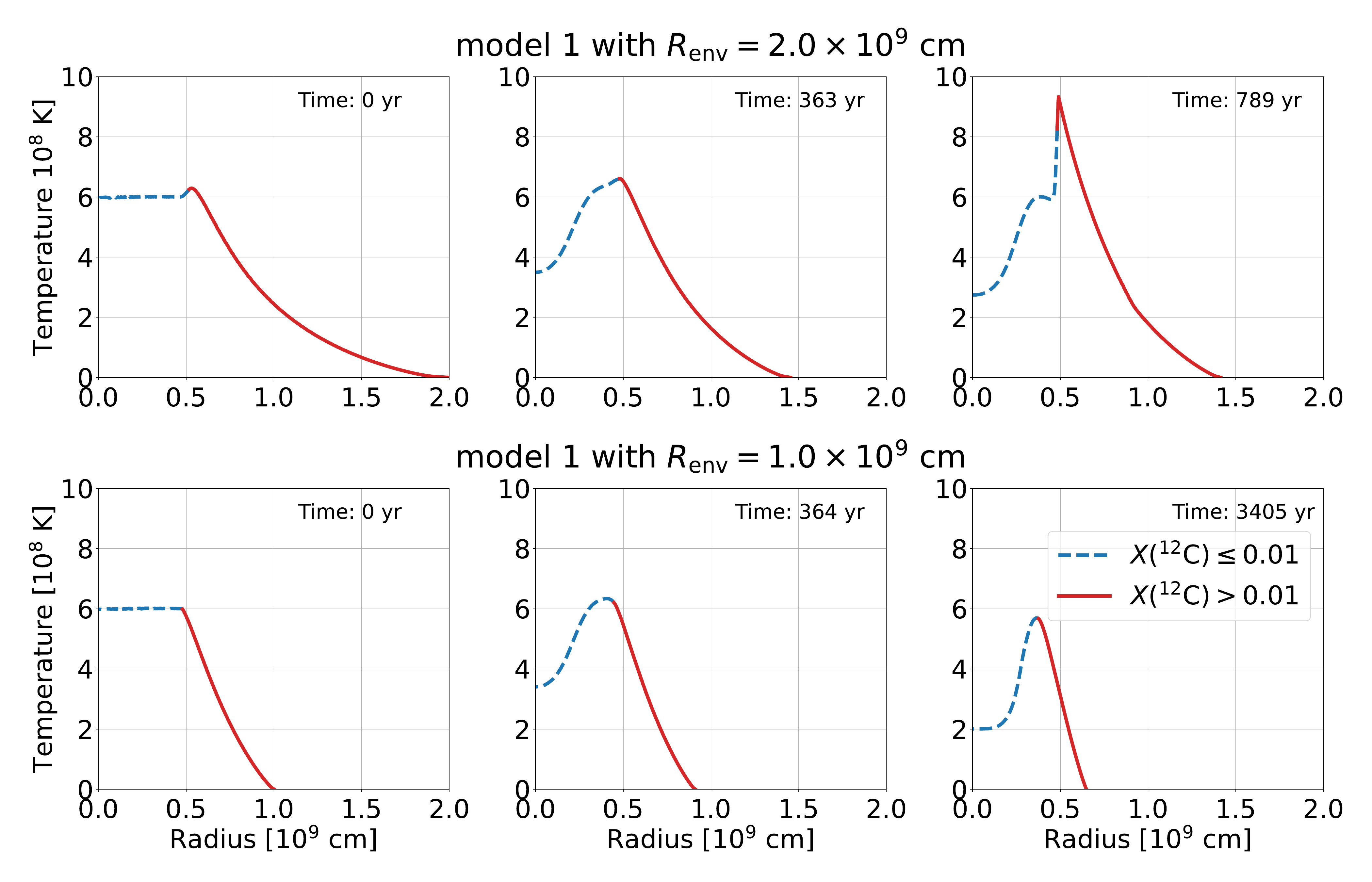}
\caption{Time evolution of the temperature profile for an ONe WD with a CO envelope. Model~1 with $R_{\rm env}=2.0\times10^9~\cm$ (upper panels, ignition case) and $R_{\rm env}=1.0\times10^9~\cm$ (lower panels, non-ignition case). The ONe WD and CO envelope are distinguished in blue and red, respectively (dashed blue portions: regions with $X(^{12}\mathrm{C})\le 0.01$; solid red portions: $X(^{12}\mathrm{C})>0.01$).}
 \label{fig:wd_evo}
\end{figure*}

Figure~\ref{fig:wd_evo} presents representative snapshots of the temperature distribution: the upper panels show the carbon ignition case (model~1 with $R_{\rm env}=2.0\times10^9~\mathrm{cm}$), and the bottom panels show a non-ignition case (model~1 with $R_{\rm env}=1.0\times10^9~\mathrm{cm}$). The ONe WD and the CO envelope are indicated by the blue dashed and red solid segments of the curves, respectively.

As soon as we start the simulation, the central parts of the WD drops in temperature due to neutrino cooling, whereas the outer parts around the core-envelope interface increases in temperature due to gravitational contraction and heat transport from the inner layers (middle panels). Depending on the model, the heat generated by gravitational contraction is supplied to the bottom of the envelope. If the supplied heat is much larger than the energy loss by transport to the cooler, more outer envelope layers, the temperature at the base of the envelope rises and ignites carbon burning. This creates a strong spike in the temperature distribution (top-right panel). In contrast, if the released gravitational energy is efficiently transported outward, the temperature does not increase, never reaches the ignition temperature (right bottom panel), and continues to cool into a fully degenerate hybrid CONe WD. This indicates that, even at fixed envelope mass, whether ignition occurs depends sensitively on the envelope thermal profile, which is here parameterized by its radial extent $R_{\rm env}$. The other models not shown here follow qualitatively similar temperature evolution to one of these representative models. Here, note that, in our calculation, although the ONe interior is set as isothermal, our models permit a small amount of heat to diffuse into the outer envelope. Such leakage is reported in other WD calculations~(e.g.,~\cite{2007MNRAS.380..933Y}). Therefore, the temperature at the core-envelope interface does not necessarily correspond to the highest temperature of the envelope.

\begin{figure*}

 \includegraphics[width=\linewidth]{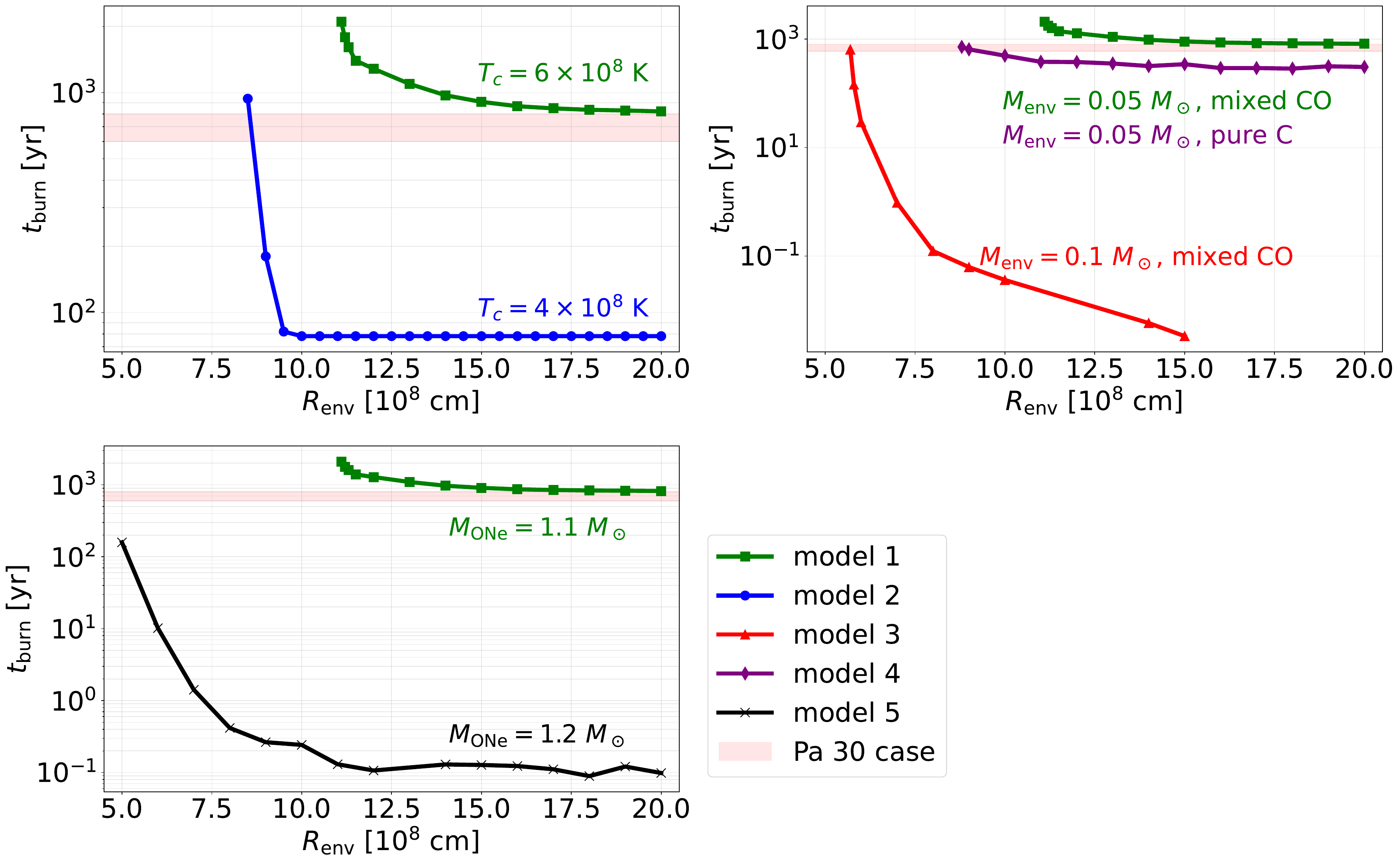}
\caption{Time to carbon ignition at the surface of ONe WDs ($t_{\rm burn}$) as a function of the initial envelope expansion radius. Each panel compares different models by varying one of the key parameters: the ONe WD mass ($M_\mathrm{ONe}$), central temperature ($T_c$), and the carbon envelope mass ($M_\mathrm{env}$). The red shaded regions show the range for the delayed wind in Pa 30: although \citet{2024ApJ...969..116K} estimates the delayed time to be $\sim810-830$~yr, we conservatively highlight 600-800 year to allow for systematic uncertainties. These bands are for visual guides only.}

 \label{fig:time_ignition}
\end{figure*}

\subsection{Parameter dependence of the ignition time}
The elapsed time ($t_{\rm burn}$) until carbon is ignited in the envelope for models 1–5 at each \(R_{\rm env}\) is shown in Fig.~\ref{fig:time_ignition}. Here, ignition is defined as when the nuclear luminosity first exceeds $10^7~L_\odot$, which is well above the Eddington luminosity of the WD ($\sim 10^5L_\odot$). The ignition delay increases systematically with decreasing \(R_{\rm env}\). This behavior reflects the steeper temperature gradient that develops in more compact envelopes: heat released at the bottom by gravitational contraction is transported outward more efficiently, which suppresses the rise of the peak temperature. Consequently, larger amounts of gravitational energy must be released to attain the ignition temperature, resulting in a smaller ignition radius \(R_{\rm bottom}(t=t_{\rm burn})\). {Here, \(R_{\rm bottom}(t)\) refers to the radius at the base of the envelope at time $t$, where $t$ is the time since the beginning of our simulation. This is consistent with Fig.~\ref{fig:env_evo_model1}, which shows the time evolution of the bottom radius of CO envelope in model~1 for $R_{\rm env} = 1.0,1.5,$ and $2.0 \times 10^9~\cm$.

\begin{figure}
 \includegraphics[width=\linewidth]{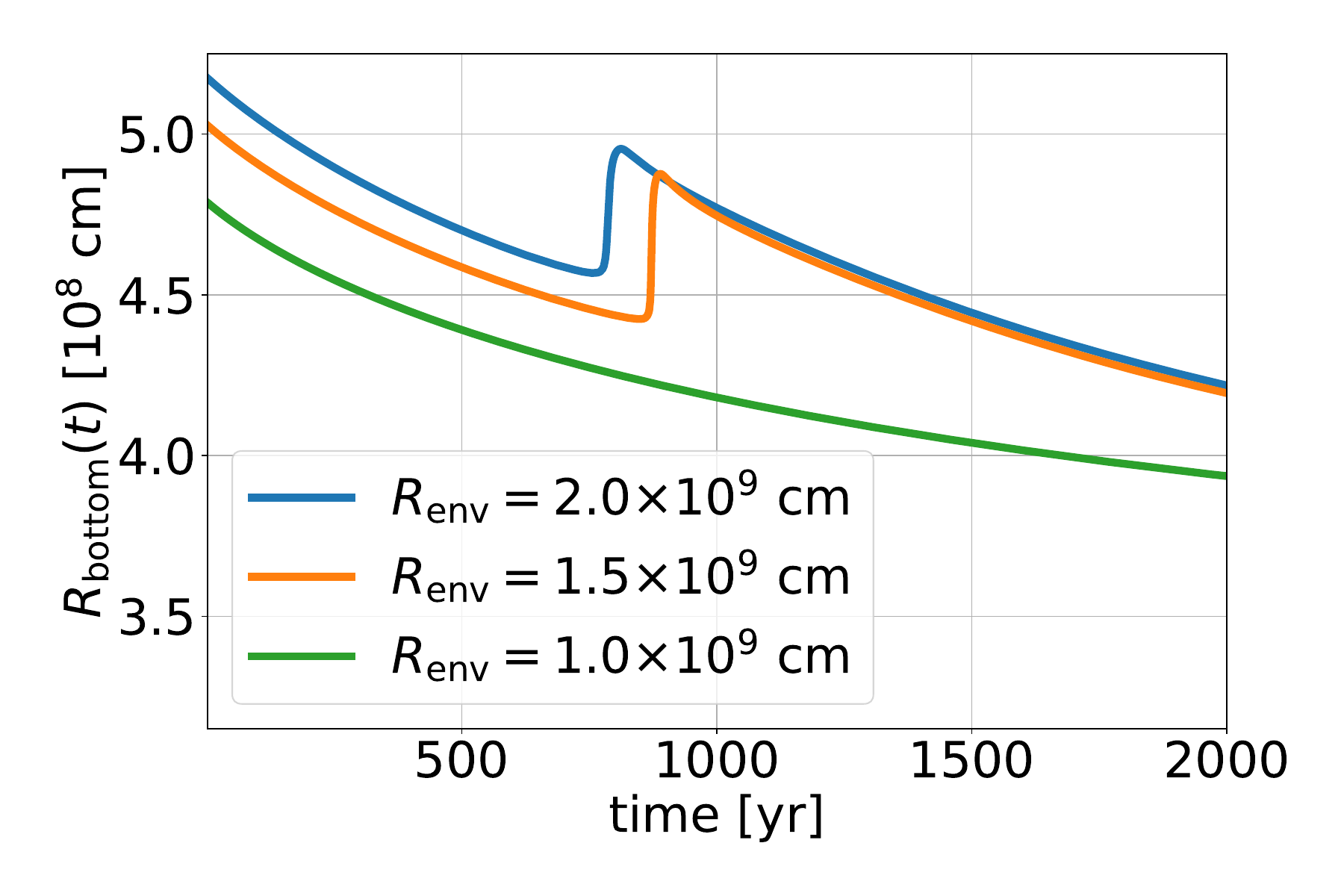}
\caption{Time evolution of bottom of the CO envelope ($R_{\rm bottom}(t)$) for model 1 with $R_{\rm env} = 1,1.5,$ and $2 \times 10^9~\cm$. As $R_{\rm env}$ decreases, the ignition timescale increases; once $R_{\rm env}$ falls below the critical threshold, ignition fails and the envelope cools and contracts.}

 \label{fig:env_evo_model1}
\end{figure}
When outward thermal losses exceed compressional heating, the peak temperature does not reach the ignition threshold and the model cools without ignition, as in the non-ignition cases discussed in Section~\ref{sec:evolve}. Non-ignition cases are omitted from the figure. The findings imply that cooler and more compact envelopes prevent the igition of carbon burning.

In this study, we adopt model~1 as the fiducial case. The upper-left panel compares models~1 and~2 to isolate the effect of $T_c$ on the ignition timescale. A higher $T_c$ results in a longer ignition timescale. A higher $T_c$ inflates the ONe component so that the bottom of the CO envelope initially lies at a larger radius $R_{\rm bottom}(t = 0)$. Because ignition requires the bottom to contract to the critical radius $R_{\rm bottom}(t = t_{\rm burn})$ where compressional heating triggers carbon burning, a larger difference between $R_{\rm bottom}(t=0)$ and $R_{\rm bottom}(t = t_{\rm burn})$ delays the ignition. This behavior can be explained with a longer thermal timescale $\tau_{\rm th}$ for hotter WDs. The upper-right panel examines variations in envelope mass (model~1 versus model~3) and composition (model~1 versus model~4). A smaller envelope mass increases the ignition timescale, again because a lighter envelope places the initial $R_{\rm bottom}(t  = 0)$ at a larger radius and therefore lengthens the contraction time $t=t_{\rm burn}$. It can also be seen that the pure-carbon case is faster to reach envelope carbon ignition than the CO-mixed case. This is because, in the pure-carbon case, the number density of carbon nuclei is higher, which enhances the $^{12}$C–$^{12}$C reaction rate and thereby lowers the temperature required to trigger carbon ignition relative to the CO-mixed case (equivalently, the ignition radius $R_{\rm bottom}(t = t_{\rm burn})$ is larger). Nevertheless, the magnitude of this composition effect is modest compared with the differences produced by varying other parameters. The lower-left panel shows that a higher WD mass shortens the ignition time. A more massive WD has a smaller initial $R_{\rm bottom}(t=0)$ and stronger gravity, so the envelope contracts to $R_{\rm bottom}(t = t_{\rm burn})$ more rapidly. These results are consistent with Fig.~\ref{fig:env_evo_models}, which shows the time evolution of the bottom radius of the CO envelope for models~1, 2, and 4 with $R_{\rm env} = 2.0\times 10^9~\cm.$  Models~3 and 5 are omitted because their time to ignition is too short.
\begin{figure}
 \includegraphics[width=\linewidth]{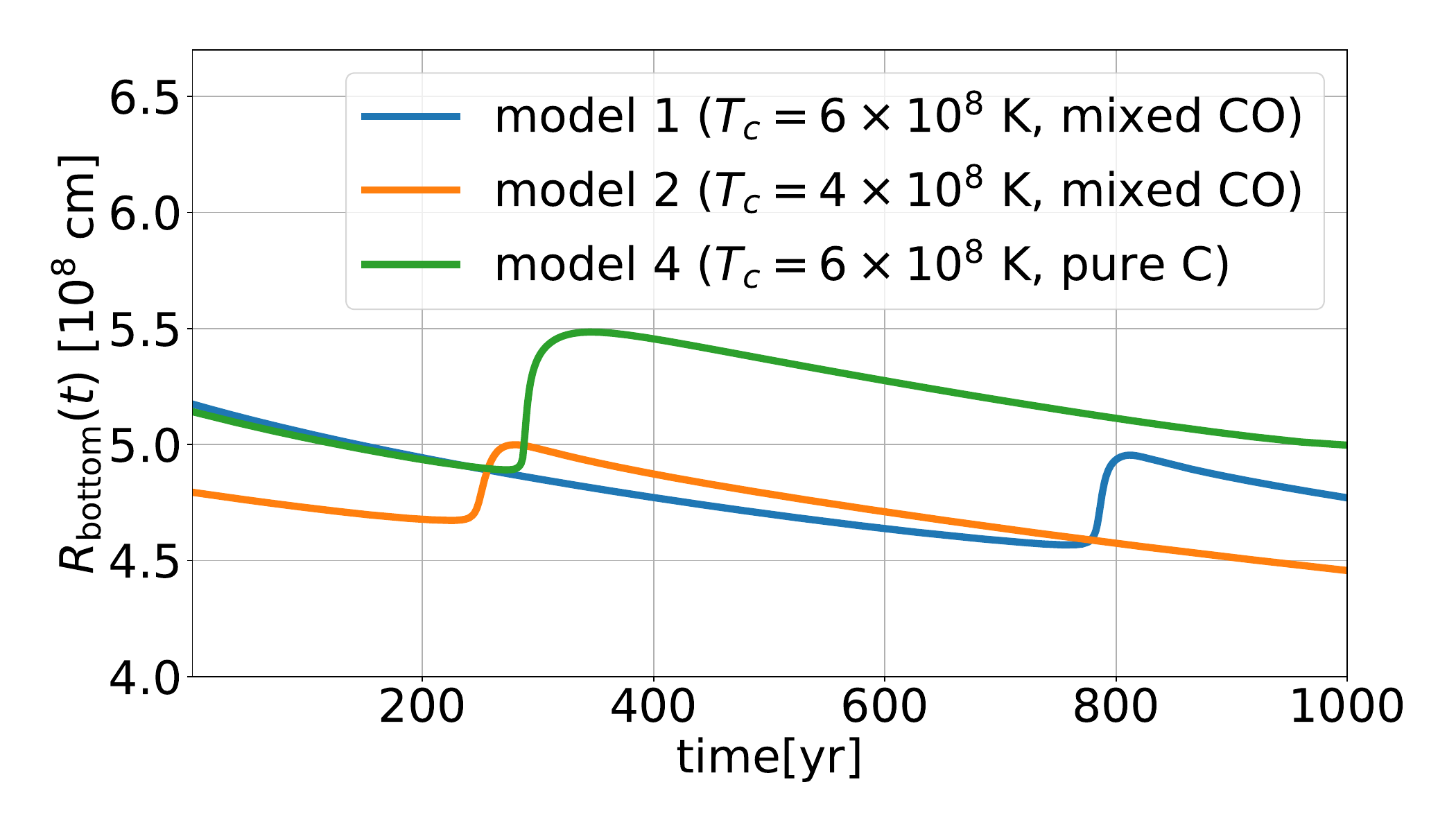}
\caption{Time evolution of the bottom radius of the CO envelope ($R_{\rm bottom}(t)$) for models 1, 2, and 4 with $R_{\rm env} = 2.0\times 10^9$ cm. Models 3 and 5 reach ignition on timescales too short to be informative at this scale and are therefore omitted.}

 \label{fig:env_evo_models}
\end{figure}

Collectively, these results indicate that long, $\sim800$~yr delayed carbon ignition can be realized across a broad parameter space provided the envelope extent, i.e., $R_{\rm env}$ is relatively small (or has low entropy). The precise necessary conditions of the envelope thermal properties that cause ~800 yr delay remain uncertain. Under a probabilistic interpretation, the conditions most likely to yield $\sim800$~yr delay are relatively high central temperature ($T_c\approx6\times10^8~\mathrm{K}$), the mass of fallback material that is not excessively large ($M_{\rm env}\approx0.05~M_\odot$), and a moderately low WD mass ($M_{\rm WD}\approx1.1~M_\odot$). The ignition timescale is also found to be sensitive to modest variations in $T_c$, $M_{\rm env}$, and $M_{\rm WD}$; even changes by factors of a few in these quantities produce order-of-magnitude shifts in $t_{\rm burn}$.
\section{Discussions}
\label{sec:discussion}
\subsection{Implication on SN 1181 Progenitor system}
As shown in Sec.~\ref{sec:results}, delayed carbon ignition on timescale of $\sim 800~\yr$ requires a relatively hot central WD. In an ONe WD + CO WD merger scenario with detonation of the CO WD, the secondary CO WD is destroyed and little heat couples to the surviving ONe WD core via CO detonation, so the central temperature does not increase. In general, the neutrino–cooling timescale for \(1.1\text{--}1.2~M_\odot\) WD at \(T_c \simeq 3\times10^8~\mathrm{K}\) is \(\sim10^5~\yr\), which is much shorter than the WD formation timescale ($10^8-10^9~\yr$), therefore, the core of the surviving ONe WD that has existed for much longer timescale than the cooling timescale and is expected to have cooled substantially, resulting in a relatively low temperature.
In contrast, in the pure-deflagration scenario the flame propagates both outward and inward into the WD: the outward branch ejects the accreted matter, forming the SN~Iax ejecta, whereas the inward branch deposits heat into the core and raises $T_c$, which could cause delayed carbon ignition at the envelope formed by the fallback material. Our results suggest that a hot central WD is required to trigger such delayed carbon ignition. This requirement, in turn, favors the pure-deflagration scenario for SN~Iax~1181. Therefore, if the delayed wind in SN~1181 indeed originates from delayed carbon ignition, the pure-deflagration interpretation is preferred.

If SN~1181 is a pure-deflagration SN~Iax, a surviving He star companion is expected, however, none has been identified within Pa~30. 
\red{Whether the surviving companion stays bound to the compact remnant depends on the amount of mass lost in the SN ejecta and the kick imparted to the remnant due to asymmetries in the explosion.}
\red{Pure-deflagration simulations still leave the kick velocity of the remnant highly uncertain~(e.g., \cite{2012ApJ...761L..23J,2014MNRAS.438.1762F}), so it is not firmly known whether the WD–He star binary will be disrupted. However, because the ejecta mass is expected to be well below half of the total system mass, the system should, in most cases, remain bound.} \red{In the context of SNe~Ia, it is known that when the WD in a WD-He star binary explodes, the He star companion can be shock-heated and subsequently expand; its radius may increase by a factor of $\sim2$ on a timescale of $\sim10^{6}$--$10^{8}\,\mathrm{yr}$~\citep{2019ApJ...887...68B}; see also \cite{2025ApJ...992..108W} for the case of a He WD companion. Such shock heating and expansion of the He-star donor can make it easier for the companion to fill its Roche lobe, allowing Roche-lobe overflow to resume and mass transfer onto the bound WD remnant. Depending on the post-explosion orbital separation and the mass-transfer rate, the interaction may remain stable and produce a long-lived accreting binary, or become unstable, particularly if the WD remnant itself is thermally inflated, leading to a common-envelope like phase and ultimately a merger into a single object. Moreover, continued He accretion onto an ONe WD remnant may trigger unstable He burning, potentially producing recurrent He-shell flashes on the ONe WD sruface~\citep{2021RAA....21...34G}. If the remnant can retain a significant fraction of the accreted material and grow in mass, it may eventually undergo accretion-induced collapse to a neutron star (e.g., \cite{1991ApJ...367L..19N}). Alternatively, if a sufficiently massive He layer accumulates on the surface, a He-shell detonation may occur, giving rise to a faint, rapidly evolving thermonuclear transient (i.e., an SN .Ia; \cite{2007ApJ...662L..95B,2010ApJ...715..767S}).}

\red{If the binary system is not disrupted,} two possibilities remain at least. First, if the WD is enshrouded by an optically thick wind with the photosphere at $\sim 10^{10}~\cm$, the He star could lie within it and be obscured. Second, the He star companion may lie outside the photosphere but be intrinsically faint—particularly at UV wavelengths—and has therefore escaped detection in existing observations. The two cases are examined in the following subsections.

\subsubsection{He star inside the photosphere}
If the He star lies inside the wind photosphere, then with the photospheric radius of $\sim10^{10}~\cm$ and the WD radius of $\sim10^{8}~\cm$, the He star radius of $\sim10^{9}~\cm$ and the separation of a few $\times10^{10}~\cm$ are plausible. In this configuration, the WD that launches the fast wind orbits its companion with a period of a few minutes, and there are sightlines where the He star—an order of magnitude larger than the WD—partially covers the wind. In addition, the WD itself moves at velocities of order $1000~\mathrm{km}~\s^{-1}$, corresponding to 5–10 $\%$ of the wind velocity, which may imprint density inhomogeneities of similar amplitude in the optically thick wind photosphere. Consequently, the optically thick wind photosphere is expected to exhibit quasi-periodic variability on minute timescales. Therefore we conducted a 1-fps video observation with Tomo-e Gozen, a wide-field CMOS camera mounted on a Kiso Schmidt telescope \citep{2018SPIE10702E..0JS}.

The light curve obtained is shown in Figure~\ref{fig:video_lc}. The central WD photosphere does not show quasi-periodic variability. In addition, we compare the central photosphere with a nearby star separating 55 arcsec away (Gaia DR3 526287601489793920). As a result, these two light curves exhibit a similar trend, which is mainly attributed to background variability. These results suggest that there is no clear variability in the light curve of the photospheric emission.

The TESS light curves show neither significant coherent periodicity between 40~seconds and 10~days, nor any long-term trend \citep{2023MNRAS.523.3885S}. The lack of variations of the photosphere means that the He star is most likely not inside the photosphere, unless the system is viewed almost pole-on.

\begin{figure*}
 \includegraphics[width=\linewidth]{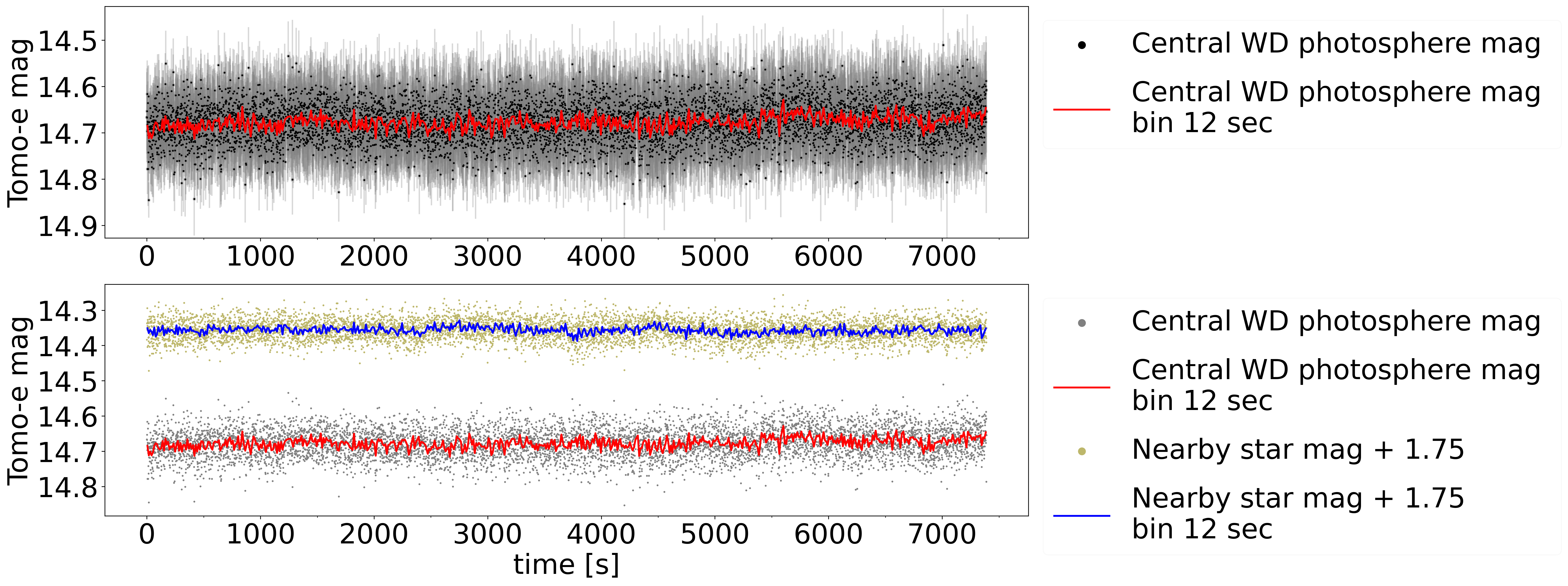}
\caption{1-fps light curve of the central WD photosphere obtained with Tomo-e Gozen at the Kiso Observatory, University of Tokyo. Top: Light curve of the central photosphere with error bars reflecting both photometric and zero-point uncertainties. Bottom: Light curves of the central photosphere and a nearby comparison star, shown without error bars. Individual measurements are plotted as yellow dots (comparison star, upper) and gray dots (central photosphere, lower); binned light curves are overplotted as blue and red curves, respectively. The x-axis is time since the start of the observation, and the y-axis is the clear-filter magnitude from Tomo-e Gozen.}

\label{fig:video_lc}
\end{figure*}
\subsubsection{He star outside the photosphere}
Because the optically thick wind photosphere exhibits no variability compatible with the orbital period, the companion He star is expected to lie outside the photosphere, if it exists. He stars are intrinsically bright at ultraviolet wavelengths; however, Hubble Space Telescope (\textit{HST}) spectroscopic observations (program GO-15864, PI: G. Gräfener) on 2020 November 4 detected no counterpart attributable to a He-rich source from the He star \citep{Lykou2022}. The spectral coverage was $1150$–$3150$~\AA, with a slit width of $0.2^{\prime\prime}$. Given that the central source radiates at $L \simeq 10^{38.4}\ \mathrm{erg\ s^{-1}}$, the companion He star could be hidden in its spectra. The ratio of the He star surface flux to that of the central photosphere is
\begin{equation}
   \mathrm{Ratio}(\lambda) =  \frac{R_\mathrm{He}^2\,B_\lambda\!\left(T_\mathrm{He}\right)}
         {R_\mathrm{photo}^2\,B_\lambda\!\left(T_\mathrm{photo}\right)},
\end{equation}
where \(R_\mathrm{He}\) and \(R_\mathrm{photo}\) are the radii of the He star and the central photosphere, and \(T_\mathrm{He}\) and \(T_\mathrm{photo}\) are their surface temperatures. Here, \(B_\lambda\) denotes the Planck spectral radiance per unit wavelength. From the past observations, \(R_\mathrm{photo}\sim10^{10}~\cm\) and \(T_\mathrm{photo}\sim200,000~\mathrm{K}\) are obtained~(e.g.,~\cite{2019Natur.569..684G,Lykou2022}). The ratios for He star masses of \(0.5\), \(1\), and \(2\,M_\odot\) are shown in Figure~\ref{fig:he_star}. He star radii and temperatures are estimated using the fitting formulae of \citet{2000MNRAS.315..543H}. For \(M_\mathrm{He}\!\sim\!2\,M_\odot\), the He star would dominate the spectrum, which is inconsistent with the absence of such features in current observations. In contrast, for \(M_\mathrm{He}\!\lesssim\!1\,M_\odot\) the He star contributes only at the level of a few tens of percent of the central photospheric flux, which is not inconsistent with the past observations. Since no search for hunting the He star spectral signatures have been conducted, a detailed analysis of the existing \textit{HST} data or deeper follow-up observations remain important future work.

\begin{figure}
 \includegraphics[width=\linewidth]{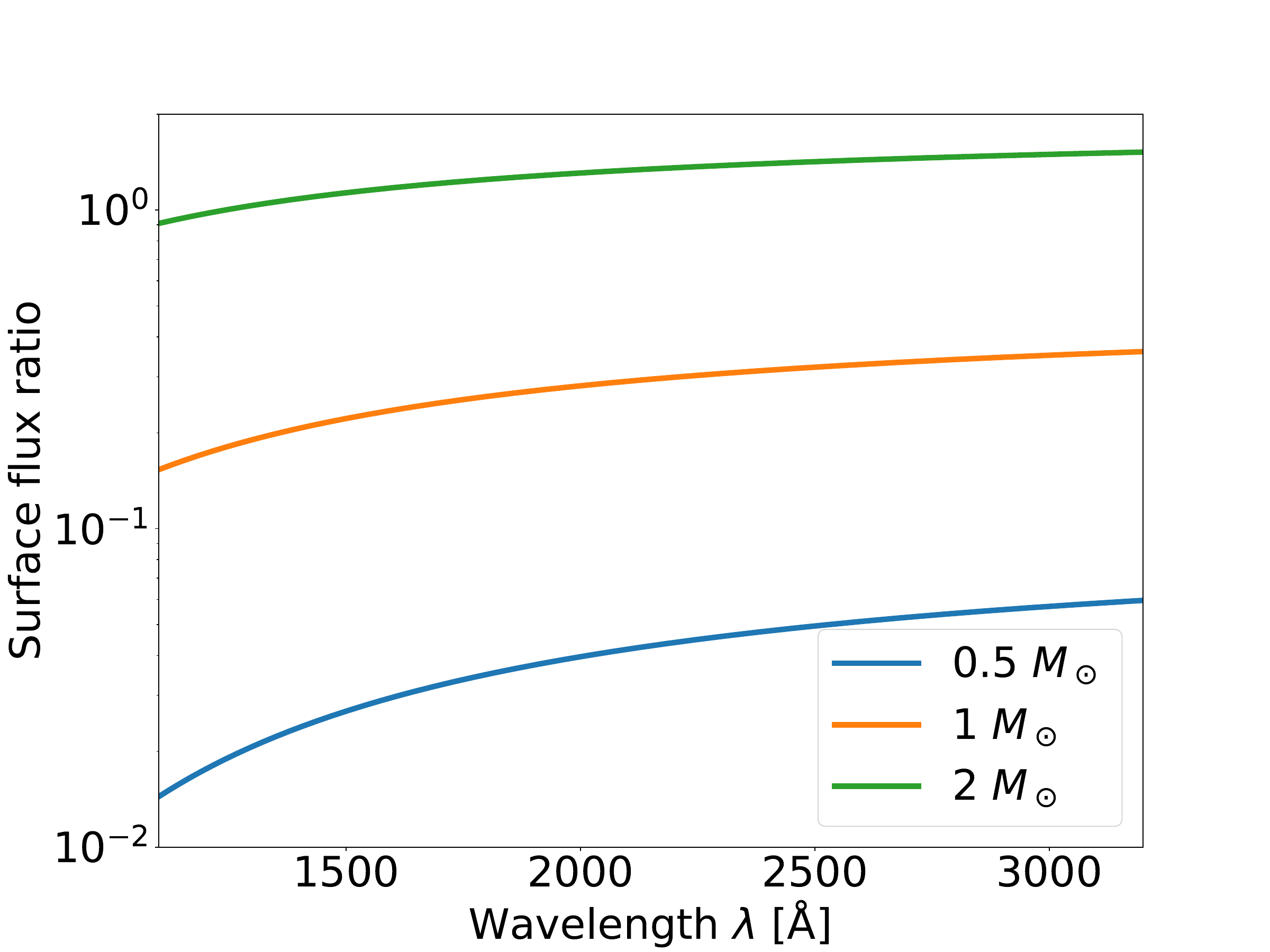}
\caption{The ratio of the surface flux of the He star to the central photosphere for He star mass $0.5~M_\odot$ (blue), $1~M_\odot$ (orange), and $2~M_\odot$ (green).}
 \label{fig:he_star}
\end{figure}

If the WD + He star binary progenitor system was disrupted by SN 1181, the surviving He star may fall outside the slit used in the past \textit{HST} spectroscopic observations (0.2$^{\prime\prime}$, corresponding to $\sim 10^{15}$–$10^{16}$~cm). A dedicated, wide-field UV imaging campaign of the central region of radius $\sim 10^{18}$~cm ($\sim 30$~arcsec) with the \textit{HST} would therefore be capable of detecting the surviving He star, should it be present. In addition, mid-infrared imaging with the James Webb Space Telescope (\textit{JWST}) under GO~9111 (Cycle~4; PI: I.~Caiazzo) is scheduled for the entire Pa~30 nebula, and thus He line emission from the surviving He star could be detected if present.

\subsection{Caveats}
In this study, we consider the fast wind is caused by the carbon ignition on the WD surface. However, the driving mechanism of the fast wind remains uncertain. While \citet{2019ApJ...887...39K} suggest that a rapidly rotating (\(\Omega \sim 0.2\text{--}0.5~\mathrm{s}^{-1}\)) and highly magnetized (\(B \sim 10^7~\mathrm{G}\)) WD can drive the wind, attaining such extreme spin is difficult to reconcile with the pure-deflagration scenario and instead favors the double WD merger scenario as an alternative origin.
However, if rapid rotation launches an equatorially enhanced wind, the resulting asphericity~\citep{2024ApJ...963...26Z} would contradict the near-spherical appearance inferred from observations. Though alternative drivers, such as line-driven acceleration, are under investigation, a unified picture has not yet emerged. Clarifying the wind mechanism is therefore a key priority in understanding the evolution of SN 1181.

Additionally, because this study does not model the energy distribution of the fallback CO material in detail, the thermal energy content of the resulting CO envelope remains unconstrained; accordingly, we treat the envelope extent $R_{\rm env}$ as a free parameter and explore a wide range of values. It is important to self-consistently follow the evolution of the bound SN ejecta and its fallback to determine the post-fallback thermal profile as future work.

We emphasize that, among the existing scenarios, the mechanism that raises the central temperature of the surviving WD is most naturally consistent with the pure-deflagration model involving a He-star companion. However, it should be noted that merely requiring a high central temperature for the remnant WD does not, by itself, rule out other scenarios.

\section{Conclusion}\label{sec:conclusion}
We use the public stellar evolution code \texttt{MESA} to follow the evolution of an ONe WD with a CO fallback envelope until carbon ignition. Motivated by SN~1181/Pa~30—where the fast wind appears to switch on only in recent years—we interpret the delay as the delayed ignition of carbon shell burning powered by SN ejecta fallback. Our models reproduce centuries-long delays provided the post-explosion WD core is hot ($T_c\approx6\times10^8~\mathrm{K}$), the fallback mass is modest ($M_{\rm env}\approx0.05~M_\odot$), and the WD mass is moderately low ($M_{\rm WD}\approx1.1~M_\odot$). The ignition timescale is highly sensitive to these parameters.

The hot surviving WD—consistent with other SNe Iax—supports a pure-deflagration origin for SN 1181, implying a WD + He star progenitor. We therefore search for the surviving He star. If embedded within the optically thick wind from the WD, the companion would be photometrically hidden but should imprint short-timescale variability on the wind photosphere. Therefore we conducted 1-fps Tomo-e Gozen monitoring observation, but no such variability was detected, indicating that the companion lies outside the photosphere. Yet archival \textit{HST} spectroscopy does not reveal the companion, which requires the He star to be faint enough for being veiled by the central source, favoring a mass $\lesssim 1~M_\odot$. In other cases, the progenitor binary could have been disrupted by the type Iax SN 1181 explosion such that the companion has moved beyond the previously observed \textit{HST} field. In this case, wide-field \textit{HST} or \textit{JWST} imaging would offer the best prospects.
\begin{ack}
 Takatoshi Ko thanks Daichi Tsuna for a fruitful discussion. Takatoshi Ko is supported by JSPS KAKENHI grant Number 24KJ0672. This work was made possible through the support of the Enrico Fermi Fellowships led by the Center for Spacetime and the Quantum, and supported by Grant ID \#63132 from the John Templeton Foundation. The opinions expressed in this publication are those of the author(s) and do not necessarily reflect the views of the John Templeton Foundation or those of the Center for Spacetime and the Quantum (Takatoshi Ko, Toshikazu Shigeyama). Taiga Sasaoka is supported by JST SPRING, Grant Number JPMJSP2108, and by FoPM, WINGS Program of the University of Tokyo. Tomo-e Gozen 1-fps video light curves were obtained using a photometry pipeline developed by Shodai Nezu at the University of Tokyo. Ryosuke Hirai acknowledges support from the RIKEN Special Postdoctoral Researcher Program for
junior scientists. Toshikazu Shigeyama acknowledges support by JSPS KAKENHI Grant Numbers JP22K03688, JP22K03671, and JP20H05639.
\end{ack}

 \section*{Data Availability}
All data underlying this article are available on reasonable request to the corresponding author.
\bibliographystyle{apj} 
\bibliography{CSM}

\end{document}